\documentstyle[11pt]{article}
\begin{document}
\newcommand{\vsp}{\vspace*{3mm}}
\newcommand{\be}{\begin{equation}}
\newcommand{\ee}{\end{equation}}
\newcommand{\bd}{\begin{displaymath}}
\newcommand{\ed}{\end{displaymath}}
\newcommand{\sgn}{~{\rm sgn}}
\newcommand{\extr}{~{\rm extr}}
\newcommand{\Equiv}{\Longleftrightarrow}
\newcommand{\prim}{~\prime}
\newcommand{\notexists}{\exists\hspace*{-2mm}/}
\newcommand{\bra}{\langle}
\newcommand{\ket}{\rangle}
\newcommand{\order}{{\cal O}}
\newcommand{\minus}{\!-\!}
\newcommand{\plus}{\!+\!}
\newcommand{\erf}{{\rm erf}}
\newcommand{\bm}{\mbox{\boldmath $m$}}
\newcommand{\br}{\mbox{\boldmath $r$}}
\newcommand{\bq}{\mbox{\boldmath $q$}}
\newcommand{\hbm}{\hat{\mbox{\boldmath $m$}}}
\newcommand{\hbr}{\hat{\mbox{\boldmath $r$}}}
\newcommand{\hbq}{\hat{\mbox{\boldmath $q$}}}
\newcommand{\bsigma}{\mbox{\boldmath $\sigma$}}
\newcommand{\bomega}{\mbox{\boldmath $\Omega$}}
\newcommand{\bphi}{\mbox{\boldmath $\Phi$}}
\newcommand{\bpsi}{\mbox{\boldmath $\psi$}}
\newcommand{\bdelta}{\mbox{\boldmath $\Delta$}}
\newcommand{\btheta}{\mbox{\boldmath $\theta$}}
\newcommand{\bxi}{\mbox{\boldmath $\xi$}}
\newcommand{\bmu}{\mbox{\boldmath $\mu$}}
\newcommand{\brho}{\mbox{\boldmath $\rho$}}
\newcommand{\bEta}{\mbox{\boldmath $\eta$}}
\newcommand{\req}{r_{\rm eq}}

\title{\bf Order Parameter Flow in the SK
Spin-Glass I: Replica Symmetry}
\author{A.C.C. Coolen \and D. Sherrington}
\date{}
\maketitle
\begin{center}
Department of Physics - Theoretical Physics\\
University of Oxford\\
1 Keble Road, Oxford OX1 3NP, U.K.
\end{center}
\vsp
\begin{center} PACS: 75.10N, 05.20 \end{center}
\vsp
\begin{abstract}
We present a theory to describe
the dynamics of the Sherrington-Kirkpatrick spin-glass with
(sequential)
Glauber
dynamics in terms of
deterministic flow equations for macroscopic parameters.
Two transparent assumptions allow us to close the macroscopic
laws. Replica theory enters as a tool in the calculation of the time-dependent
local field distribution.
The theory produces  in a natural way
dynamical generalisations of the AT- and zero-entropy lines and of
Parisi's order parameter function $P(q)$.  In equilibrium we recover
the standard results from equilibrium statistical mechanics.
In this paper we make the replica-symmetric ansatz, as
a first step towards calculating the order parameter flow.
Numerical simulations support our assumptions and suggest that our
equations describe the shape of the local field distribution and
the macroscopic dynamics reasonably well in the region where
replica symmetry is stable.
\end{abstract}

\pagebreak\tableofcontents

\pagebreak\section{Introduction}

The Sherrington-Kirkpatrick (SK) spin-glass model \cite{SK} describes
a collection of $N$ Ising spins, coupled by exchange interactions which
are drawn at random from a Gaussian distribution. These interactions
represent quenched (frozen) disorder.
The equilibrium statistical mechanical description  of the
SK model  seems to have
reached a stable fixed-point, built on replica theory with,
at least in
the spin-glass phase,
broken replica symmetry a la Parisi \cite{parisi}. A clear and
extensive description
of the formalism developed since 1975
and most of the relevant references  can be found in textbooks like
the ones by Mezard et al \cite{mezardetal}  and Fisher and Hertz
\cite{fisherhertz}.

With respect to the dynamical properties of the SK model, the situation
seems different. The early dynamical studies, like \cite{EA,KF,KS},
were more or less of a pilot character, employing mean
field approximations (MFA) and linearisations of the exact dynamic ensemble
averages.
Analytical work beyond MFA published so far
has mostly concentrated
on Langevin dynamics for soft spins, as opposed to Ising spins
\cite{sompolinsky,SZ,horner,cugliandolokurchan}. In
the Langevin case the standard procedure (described in detail in e.g.
\cite{zinnjustin} and \cite{fisherhertz}) is to construct a generating
functional from a path integral representation of the microscopic
state probability, which can subsequently be averaged over the
quenched disorder (i.e. the random exchange interactions). This
leads to a saddle-point problem, the limit
$N\rightarrow\infty$ can be taken and one obtains a complicated set of
equations for correlation- and response functions. These can
be interpreted in terms of a Langevin equation for a single spin with a
retarded
self-interaction and a noise term with non-trivial moments.
In order to proceed from this stage, additional assumptions,
restrictions  or
approximations are needed, like expansions near critical lines or near
equilibrium. By construction, in these theories only timescales which do not
diverge with
$N$ are described.
The approach followed by Sompolinsky in \cite{sompolinsky}
is different: here a hierarchy
of time-scales is introduced, all of which diverge for
$N\rightarrow\infty$, but in a strict order.
The case of Glauber \cite{glauber} dynamics for Ising spins was studied by
Sommers
\cite{sommers}, who developed a path integral formalism by performing
manipulations on the solution (in the form of a time-ordered product)
of the master equation. His method, although subsequently applied by
other
authors to related models like the non-symmetric SK
model \cite{riegeretal}, was later criticised by Lusakowski
\cite{lusakowski}. As far as we are aware, the issue of the correctness
(or otherwise) of the Sommers approach has not been
settled.

At zero temperature the SK model shows strong remanence effects (see
e.g. Kinzel \cite{kinzel}),
with a non-exponential decay of the magnetisation. Only recently
numerical evidence has been published \cite{cugliandoloetal} which
suggests that infinite-range models such as the SK model even exhibit
ageing effects of the type observed in experiments on real
spin-glasses \cite{fisherhertz,vincentetal}, which until now were
always assumed to be typical for finite-range models and therefore
 explained using scaling arguments for growing domains.

Motivated by the non-trivial dynamical phenomena exhibited by
the SK model and by the restricted
theoretical understanding  of the Glauber dynamics (as
opposed to the continuous Langevin approach), we develop
in this paper a theory to describe
the Glauber dynamics of the SK model in terms of
deterministic flow equations for two macroscopic state variables: the
magnetisation $m$ and the spin-glass contribution $r$ to the energy. Our
reasons for choosing these two quantities as dynamic order
parameters, in favour of a dynamical equivalent of
the spin-glass order parameter $q$ or its distribution $P(q)$, are:
\begin{enumerate}
\item On finite time-scales both $m$ and $r$ evolve in time deterministically
in
the limit $N\rightarrow\infty$.
\item The Hamiltonian of the SK model can be expressed solely in terms of $m$
and $r$.
\item In thermal equilibrium $m$ and $r$ are self-averaging with respect to the
quenched disorder in the
limit $N\rightarrow\infty$ (since $m$ and the free
energy are \cite{mezardvirasoro}).
\item Both $m$ and $r$ are instantaneous functions of time for a single
system, whereas $P(q)$ involves correlations between different times
or systems.
\end{enumerate}
The key to closing the deterministic laws is to calculate the distribution
of
time-dependent local aligment fields. Two transparent physical
assumptions allow us to calculate this distribution analytically and find a
{\em closed}
set of flow equations for our two order parameters.
The theory produces  in a natural way
dynamical generalisations of the AT- and zero-entropy lines and of
Parisi's order parameter function $P(q)$. In equilibrium we recover
the standard results from equilibrium statistical mechanics.
The present formalism has previously been applied succesfully to a
related model: the Hopfield neural network model near saturation
\cite{coolensherrington}.

In our view the main appeal of our formalism is its
transparency. The theory is formulated in terms of two directly
observable macroscopic state variables and, apart from two
simple assumptions, derived directly from
the microscopic stochastic equations. Secondly, an interesting
difference with existing approaches is the way in which replica
theory enters. In the standard Langevin approach (after having taken
the limit $N\rightarrow\infty$) one ends up with quantities and
equations
very much like the ones encountered in equilibrium replica theory,
with replica indices replaced by time arguments.
In Sompolinsky's theory replica indices are replaced by labels of
the hierarchy of time-scales. In contrast, in the present formalism
replica theory
enters as a mathematical tool in calculating the time-dependent
distribution of local alignment fields.
The only uncertainty in the status of
the theory originates from the two closure assumptions, since all
subsequent calculations can in principle be performed exactly.
Both are supported to a certain
extent by evidence from numerical
simulations. A recent study of an exactly solvable toy model
\cite{CF}, stimulated by the work reported here and in
\cite{coolensherrington},
suggests that the proposed closure procedure succeeds
in capturing the
main physics in a closed set of transparent
deterministic equations and is exact for $t=0$ and $t=\infty$, but does not
reproduce all temporal
characteristics for intermediate times.  Since the closure procedure
is based on the elimination of microscopic memory effects,
the theory can contribute to a better
understanding of the relation between the microscopic processes and
correlations and
the macroscopic measures of complexity, such as the order parameter
$P(q)$.

In this paper we develop the general formalism.  However, in
calculating the order parameter flow explicitly we will
make the replica-symmetric RS ansatz. We will show that in
most of the flow diagram replica symmetry is stable.
In the region where the RS solution is unstable
the flow {\em direction} is still described correctly
 and the RS theory even predicts non-exponential
relaxation for $T\rightarrow0$, but the RS equations fail to describe a
rigorous slowing down
which, according to simulations, sets in near the de Almeida-Thouless
\cite{AT} line.
In a subsequent paper we shall address the implications of replica symmetry
breaking.

\section{Dynamics of the Sherrington-Kirkpatrick Spin-Glass}

\subsection{Definitions and Macroscopic Laws}

The Sherrington-Kirkpatrick (SK) spin-glass model \cite{SK} describes $N$ Ising
spins
$\sigma_{i}\in\{-1,1\}$ with infinite-range exchange
interactions $J_{ij}$:
\be
J_{ij}=\frac{1}{N}J_0+\frac{1}{\sqrt{N}}J
z_{ij}~~~~~~~~(i<j)
\label{eq:interactions}
\ee
where the quantities $z_{ij}$, which represent quenched disorder,  are drawn
{\em independently} at random
from a Gaussian distribution with $\bra z_{ij}\ket=0$ and
$\bra z^2_{ij}\ket=1$.

The evolution in time of the microscopic state probability
$p_{t}(\bsigma)$ is of the Glauber \cite{glauber}
 form, described by  a continous-time master equation:
\be
\frac{d}{dt}p_{t}(\bsigma)=\sum_{k=1}^{N}\left[p_{t}(F_k\bsigma)w_{k}(F_k\bsigma)-p_{t}(\bsigma)w_{k}(\bsigma)\right]
\label{eq:master}
\ee
in which $F_k$ is a spin-flip operator
$F_{k}\Phi(\bsigma)\equiv\Phi(\sigma_1,\ldots,-\sigma_k,\ldots,\sigma_N)$ and
the transition rates $w_k(\bsigma)$ are
\bd
w_{k}(\vec{s})\equiv\frac{1}{2}\left[1-\sigma_k\tanh[\beta
h_k(\bsigma)]\right]~~~~~~~~
h_{i}(\bsigma)\equiv\sum_{j\neq i}J_{ij}\sigma_{j}+\theta
\ed
which leads to the required standard equilibrium distribution
\bd
p_{\rm eq}(\bsigma)~\sim~e^{-\beta H(\bsigma)}~~~~~~
H(\bsigma)\equiv-\sum_{i< j}\sigma_i
J_{ij}\sigma_j-\sum_i\theta_i\sigma_i
\ed
(for numerical simulations we resort to a discrete-time sequential
process, where the $w_k(\bsigma)$ are interpreted as transition
probabilities and with iteration steps of duration $1/N$. For
$N\rightarrow\infty$ this must
reproduce the physics of the continuous-time equation \cite{bedeaux}).
The energy per spin can be written in terms of two macroscopic quantities
\be
m(\bsigma)=\frac{1}{N}\sum_i\sigma_i~~~~~~~~r(\bsigma)=\frac{1}{N\sqrt{N}}\sum_{i<j}\sigma_i
z_{ij}\sigma_j
\label{eq:orderparameters}
\ee
\be
H(\bsigma)/N= -\frac{1}{2}J_0 m^2(\bsigma)-\theta m(\bsigma)
-Jr(\bsigma)+\frac{1}{2}J_0/N
\label{eq:energyperspin}
\ee
These two observables, the magnetisation and the energy contribution
induced by the quenched variables $\{z_{ij}\}$, will be
used to define a {\em macroscopic state}.
The corresponding macroscopic probability distribution is
\be
{\cal
P}_t(m,r)\equiv\sum_{\bsigma}p_t(\bsigma)~\delta\left[m-m(\bsigma)\right]
\delta\left[r-r(\bsigma)\right]
\label{eq:macroprobability}
\ee
By inserting the microscopic equation (\ref{eq:master}) and after
defining the `discrete derivatives' $\Delta_i f(\bsigma)\equiv
f(F_i\bsigma)-f(\bsigma)$, we obtain
\bd
\frac{d}{dt}{\cal P}_{t}(m,r)
=-\frac{\partial}{\partial m}\left\{P_t(m,r)\bra \sum_{i=1}^N
w_i(\bsigma)\Delta_i m(\bsigma)\ket_{m,r;t}\right\}~~~~~~~~~~~~~~~~~~~~
\ed
\be
-\frac{\partial}{\partial r}\left\{P_t(m,r)\bra \sum_{i=1}^N
w_i(\bsigma)\Delta_i r(\bsigma)\ket_{m,r;t}\right\}+\order(N\Delta^2)
\label{eq:macromaster}
\ee
with the sub-shell average
\bd
\bra \Phi(\bsigma)\ket_{m,r;t}\equiv \frac{
\sum_{\bsigma}p_t(\bsigma)\delta\left[m\minus
m(\bsigma)\right]\delta\left[r\minus r(\bsigma)\right]\Phi(\bsigma)
}{
\sum_{\bsigma}p_t(\bsigma)\delta\left[m\minus
m(\bsigma)\right]\delta\left[r\minus r(\bsigma)\right]
}
\ed
The local alignment fields and the `discrete derivatives' are given by
\bd
h_i(\bsigma)=J_0 m(\bsigma)+ J z_i(\bsigma)+\theta+\order(\frac{1}{N})
{}~~~~~~~~
z_i(\bsigma)\equiv\frac{1}{\sqrt{N}}\sum_{j\neq i}z_{ij}\sigma_j
\ed
\bd
\Delta_i m(\bsigma)=-\frac{2}{N}\sigma_i~~~~~~~~
\Delta_i r(\bsigma)=-\frac{2}{N}\sigma_i z_i(\bsigma)
\ed
With these expressions and
the transition rates (\ref{eq:master}) we can  evaluate
(\ref{eq:macromaster}):
\bd
\frac{d}{dt}{\cal P}_{t}(m,r)
=-\frac{\partial}{\partial m}\left\{P_t(m,r)\left[\bra \frac{1}{N}\sum_{i=1}^N
\tanh\beta\left(J_0 m\plus J z_i(\bsigma)\plus\theta\right)
\ket_{m,r;t}-m\right]\right\}
\ed
\be
-\frac{\partial}{\partial r}\left\{P_t(m,r)\left[\bra \frac{1}{N}\sum_{i=1}^N
z_i(\bsigma)\tanh\beta\left(J_0 m\plus J z_i(\bsigma)\plus\theta\right)
\ket_{m,r;t}-2r\right]\right\}+\order(\frac{1}{N})
\label{eq:secondmacromaster}
\ee
In the limit $N\rightarrow\infty$ equation
(\ref{eq:secondmacromaster}) acquires the Liouville form and describes
deterministic flow at the macroscopic level $(m,r)$.
The evolution of the dynamic order parameters  $(m,r)$ is governed by the
flow equations
\be
\frac{d}{dt}m=\int\!dz~D_{m,r;t}[z]\tanh\beta\left(J_0 m+Jz+\theta\right) -m
\label{eq:mequation}
\ee
\be
\frac{d}{dt}r=\int\!dz~D_{m,r;t}[z] z\tanh\beta\left(J_0 m+Jz+\theta\right) -2r
\label{eq:requation}
\ee
All complicated terms are concentrated in the distribution of
spin-glass contributions $z_i(\bsigma)$ to the local fields:
\be
D_{m,r;t}[z]\equiv\lim_{N\rightarrow\infty}
\frac{\sum_{\bsigma}p_{t}(\bsigma)\delta\left[m\minus
m(\bsigma)\right]\delta\left[r\minus
r(\bsigma)\right]\frac{1}{N}\sum_i\delta\left[z-z_i(\bsigma)\right]}
{\sum_{\bsigma}p_{t}(\bsigma)\delta\left[m\minus
m(\bsigma)\right]\delta\left[r\minus r(\bsigma)\right]}
\label{eq:exactnoisedist}
\ee
Thus far no approximations have been used; equations
(\ref{eq:mequation},\ref{eq:requation}) are exact for
$N\rightarrow\infty$.

\subsection{Closure of the Macroscopic Laws}

The flow equations are not yet closed:
they contain the distribution $D_{m,r;t}[z]$
(\ref{eq:exactnoisedist}), which is defined in terms of the solution
$p_t(\bsigma)$ of the microscopic equation (\ref{eq:master}).
In order to close the set
(\ref{eq:mequation},\ref{eq:requation})
we make
two simple assumptions on the asymptotic ($N\rightarrow\infty$) form
of the local field distribution $D_{m,r;t}[z]$:

\begin{figure}
\vspace*{12cm}
\hbox to \hsize{\hspace*{-1cm}\includegraphics{ps/selfav0.ps}\hspace*{1cm}}
\vspace*{-1cm}
\caption{Trajectories in the $(m,r)$ plane obtained by
performing sequential simulations of the SK model with $T=0.1$,
$J=1$ and $J_0=0$, for $t\leq 10$ iterations/spin.}
\label{fig:selfav0}
\end{figure}

\begin{figure}
\vspace*{12cm}
\hbox to \hsize{\hspace*{-1cm}\includegraphics{ps/selfav1.ps}\hspace*{1cm}}
\vspace*{-1cm}
\caption{Trajectories in the $(m,r)$ plane obtained by
performing sequential simulations of the SK model with $T=0.1$,
$J=1$ and $J_0=1$, for $t\leq 10$ iterations/spin.}
\label{fig:selfav1}
\end{figure}

\begin{figure}
\vspace*{12cm}
\hbox to \hsize{\hspace*{-1cm}\includegraphics{ps/selfav2.ps}\hspace*{1cm}}
\vspace*{-1cm}
\caption{Trajectories in the $(m,r)$ plane obtained by
performing sequential simulations of the SK model with $T=0.1$,
$J=1$ and $J_0=2$, for $t\leq 10$ iterations/spin.}
\label{fig:selfav2}
\end{figure}

\begin{description}
\item[$~(i)~$] The deterministic laws describing the evolution in time
of the order parameters $(m,r)$ are {\em self-averaging} with respect
to the distribution of the quenched contributions $z_{ij}$ to the
exchange interactions.
Therefore the local field distribution $D_{m,r;t}[z]$ is self-averaging as
well.
\item[$(ii)~$] In view of $(i)$ we assume that,  as far as the calculation of
$D_{m,r;t}[z]$ is concerned, we may assume equipartitioning of probability in
the macroscopic $(m,r)$
subshells of the ensemble.
\end{description}
Assumption $(i)$ allows us to simplify the problem by performing an
average over the (quenched) random variables $\{z_{ij}\}$.
As a consequence of assumption $(ii)$ the explicit time-dependence in
the flow equations (\ref{eq:mequation},\ref{eq:requation}) and the dependence
on
microscopic initial conditions are removed, since the
distribution  $D_{m,r;t}[z]$ will be replaced by:
\be
D_{m,r}[z]\equiv\lim_{N\rightarrow\infty}\bra
\frac{\sum_{\bsigma}\delta\left[m\minus
m(\bsigma)\right]\delta\left[r\minus
r(\bsigma)\right]\frac{1}{N}\sum_i\delta\left[z-z_i(\bsigma)\right]}
{\sum_{\bsigma}\delta\left[m\minus
m(\bsigma)\right]\delta\left[r\minus r(\bsigma)\right]}\ket_{\{z_{ij}\}}
\label{eq:noisedist}
\ee
For sequential dynamics, the first of our two assumptions is clearly supported
by experimental
evidence (sequential simulations at $T=0.1$), which we present in figures
\ref{fig:selfav0} (for
$J_0=0$, where the system evolves towards a true spin-glass state),
\ref{fig:selfav1} (for $J_0=1$, which marks the onset of a non-zero
equilibrium magnetisation) and \ref{fig:selfav2} (for $J_0=2$, where the system
evolves towards a ferro-magnetic state). Each of the flow graphs
corresponds to one particular realisation of the quenched disorder
$\{z_{ij}\}$. The initial states generating the different
trajectories (labelled by $\ell=0,\ldots,10$) were drawn
at random according to
$p_0(\bsigma)\equiv\prod_i\left[\frac{1}{2}[1\plus\frac{\ell}{10}]\delta_{\sigma_i,1}+\frac{1}{2}[1\minus\frac{\ell}{10}]\delta_{s_i,-1}\right]$,
such that
that $\bra m\ket_{t=0}=0.1\ell$ and $\bra
r\ket_{t=0}=0$.
With increasing system size, fluctuations in individual
trajectories eventually vanish and well-defined flow lines emerge, which no
longer
depend on the disorder realisation.
The second closure assumption can only be tested in such a direct manner by
comparing the actual local field distribution, measured during
simulations, with the result of evaluating (\ref{eq:noisedist}).
This will be done in a subsequent section.

In equilibrium studies the above two assumptions are in fact the basic building
blocks of analysis as well, where $(i)$ is assumed and $(ii)$ is a
consequence of the Boltzmann form of the microscopic equilibrium
distribution. Our aim is to calculate analytically the
$N\rightarrow\infty$ flow illustrated in figures \ref{fig:selfav0} to
\ref{fig:selfav2}, by combining equations
(\ref{eq:mequation},\ref{eq:requation}) with (\ref{eq:noisedist}). The
distribution (\ref{eq:noisedist}) will be calculated
using the replica method.

\subsection{The Local Field Distribution}

We use the following replica expression for writing expectation
values of a given state variable $\Phi$ over a given measure $W$:
\bd
\bra\Phi(\bsigma)\ket_W\equiv\frac{\bra\Phi(\bsigma)W(\bsigma)\ket_{\bsigma}}
{\bra W(\bsigma)\ket_{\bsigma}}=
\lim_{n\rightarrow0}\bra\Phi(\bsigma^{1})\prod_{\alpha=1}^{n}
W(\bsigma^{\alpha})\ket_{\{\bsigma^{\alpha}\}}
\ed
which allows us to write (\ref{eq:noisedist}) in the replica form.
By writing the delta-functions in
integral representation we obtain
\bd
D_{m,r}[z]=\int\frac{dx}{2\pi}e^{ixz}
\lim_{n\rightarrow0}\left[\frac{N}{2\pi}\right]^{2n}\times
{}~~~~~~~~~~~~~~~~~~~~
\ed
\bd
{}~~~~~~~~~~~~~~~~~~~~\int\!d\hbm d\hbr~
e^{iN\sum_{\alpha}\left[r\hat{r}_{\alpha}+m\hat{m}_{\alpha}\right]}
\bra e^{-i\sum_{\alpha}\hat{m}^{\alpha}\sum_{k}\sigma_{k}^{\alpha}}
M\{\bsigma^{\alpha}\}\ket_{\{\bsigma^{\alpha},z_{ij}\}}
\ed
with
\bd
M\{\bsigma^{\alpha}\}\equiv
\bra
e^{-\frac{ix}{\sqrt{N}}\sum_{k>1}z_{1k}\sigma^1_{k}
-\frac{i}{\sqrt{N}}\sum_{\alpha}\hat{r}_{\alpha}
\sum_{k>l}z_{kl}\sigma_k^{\alpha}\sigma_l^{\alpha}}
\ket_{\{z_{ij}\}}
\ed
We now perform the average over the quenched variables $\{z_{ij}\}$
in $M\{\bsigma^{\alpha}\}$, with the result:
\bd
M\{\bsigma^{\alpha}\}=
e^{-\frac{1}{2}x^2-\frac{1}{4}N\sum_{\alpha\beta}\hat{r}_{\alpha}q^2_{\alpha\beta}(\bsigma)\hat{r}_{\beta}+\frac{1}{4}\left[\sum_{\alpha}\hat{r}_{\alpha}\right]^2-x\sum_{\alpha}\hat{r}_{\alpha}q_{1\alpha}(\bsigma)\sigma_{1}^{\alpha}+\order(1/N)}
\ed
in which we have introduced the familiar order parameters
$q_{\alpha\beta}(\bsigma)\equiv\frac{1}{N}\sum_i
\sigma_i^{\alpha}\sigma_i^{\beta}$. If we again introduce appropriate
delta-functions,
\bd
1=\int\!d\bq~\delta\left[\bq-\bq(\bsigma)\right]=
\left[\frac{N}{2\pi}\right]^{n^2}\int\!d\hbq
d\bq~e^{iN\sum_{\alpha\beta}\hat{q}_{\alpha\beta}\left[q_{\alpha\beta}-q_{\alpha\beta}(\bsigma)\right]}
\ed
we can reduce the spin-averages to single-site ones. The result can
than be written in terms of an
$n$-replicated Ising spin $(\sigma_1,\ldots,\sigma_n)$:
\bd
D_{m,r}[z]=\int\frac{dx}{2\pi}e^{-\frac{1}{2}x^2+ixz}
\lim_{n\rightarrow0}\left[\frac{N}{2\pi}\right]^{n^2+2n}\int\!d\hbm d\hbr d\hbq
d\bq~e^{\frac{1}{4}\left[\sum_{\alpha}\hat{r}_{\alpha}\right]^2+\order(1/N)}
\ed
\bd
\times e^{N\Psi(\hbm,\hbr,\hbq,\bq)}~
\frac{\bra
e^{-\sum_{\alpha}\sigma_{\alpha}\left[x\hat{r}_{\alpha}q_{1\alpha}+i\hat{m}_{\alpha}\right]-i\sum_{\alpha\beta}\hat{q}_{\alpha\beta}\sigma_{\alpha}\sigma_{\beta}}\ket_{\bsigma}}
{\bra
e^{-i\sum_{\alpha}\hat{m}_{\alpha}\sigma_{\alpha}-i\sum_{\alpha\beta}\hat{q}_{\alpha\beta}\sigma_{\alpha}\sigma_{\beta}}\ket_{\bsigma}}
\ed
\bd
\Psi(\hbm,\hbr,\hbq,\bq)=
i\sum_{\alpha}\left[r\hat{r}_{\alpha}+m\hat{m}_{\alpha}\right]+i\sum_{\alpha\beta}\hat{q}_{\alpha\beta}q_{\alpha\beta}-\frac{1}{4}\sum_{\alpha\beta}\hat{r}_{\alpha}q^2_{\alpha\beta}\hat{r}_{\beta}
\ed
\be
+\log\bra
e^{-i\sum_{\alpha}\hat{m}_{\alpha}\sigma_{\alpha}-i\sum_{\alpha\beta}\hat{q}_{\alpha\beta}\sigma_{\alpha}\sigma_{\beta}}\ket_{\bsigma}
\label{eq:exponent}
\ee
For large $N$ the integral is evaluated by steepest descent and we
obtain
\be
D_{m,r}[z]=\int\frac{dx}{2\pi}e^{-\frac{1}{2}x^2+ixz}
\lim_{n\rightarrow0}
\frac{\bra
e^{-\sum_{\alpha}\sigma_{\alpha}\left[x\hat{r}_{\alpha}q_{1\alpha}+i\hat{m}_{\alpha}\right]-i\sum_{\alpha\beta}\hat{q}_{\alpha\beta}\sigma_{\alpha}\sigma_{\beta}}\ket_{\bsigma}}
{\bra
e^{-i\sum_{\alpha}\hat{m}_{\alpha}\sigma_{\alpha}-i\sum_{\alpha\beta}\hat{q}_{\alpha\beta}\sigma_{\alpha}\sigma_{\beta}}\ket_{\bsigma}}
\label{eq:replicadist}
\ee
in which the order parameters $\{\hbm,\hbr,\hbq,\bq\}$ are found by selecting
the
saddle-point of $\Psi$ (\ref{eq:exponent}),
which gives a minimum with respect to variation
of the order parameters $q_{\alpha\beta}$.
Variation of $q_{\alpha\beta}$ allows us
to eliminate already one set of conjugate parameters:
$\hat{q}_{\alpha\beta}=-\frac{1}{2}iq_{\alpha\beta}\hat{r}_{\alpha}\hat{r}_{\beta}$.
The remaining conjugate
parameters, uniquely determined by the saddle-point requirement,
turn out to be purely imaginary:
$\hat{r}_{\alpha}\equiv i\rho_{\alpha}$ and
$\hat{m}_{\alpha}\equiv i\mu_{\alpha}$, with which we obtain the following
saddle-point
equations:
\be
m=
\frac{\bra
\sigma_{\alpha}
e^{\sum_{\gamma}\mu_{\gamma}\sigma_{\gamma}+\frac{1}{2}\sum_{\gamma\delta}q_{\gamma\delta}\rho_{\gamma}\rho_{\delta}\sigma_{\gamma}\sigma_{\delta}}\ket_{\bsigma}}
{\bra
e^{\sum_{\gamma}\mu_{\gamma}\sigma_{\gamma}+\frac{1}{2}\sum_{\gamma\delta}q_{\gamma\delta}\rho_{\gamma}\rho_{\delta}\sigma_{\gamma}\sigma_{\delta}}\ket_{\bsigma}}
\label{eq:saddle_m}
\ee
\be
q_{\alpha\beta}=
\frac{\bra
\sigma_{\alpha}\sigma_{\beta}
e^{\sum_{\gamma}\mu_{\gamma}\sigma_{\gamma}+\frac{1}{2}\sum_{\gamma\delta}q_{\gamma\delta}\rho_{\gamma}\rho_{\delta}\sigma_{\gamma}\sigma_{\delta}}\ket_{\bsigma}}
{\bra
e^{\sum_{\gamma}\mu_{\gamma}\sigma_{\gamma}+\frac{1}{2}\sum_{\gamma\delta}q_{\gamma\delta}\rho_{\gamma}\rho_{\delta}\sigma_{\gamma}\sigma_{\delta}}\ket_{\bsigma}}
\label{eq:saddle_q}
\ee
\be
\sum_{\beta}q^2_{\alpha\beta}\rho_{\beta}=2r
\label{eq:saddle_r}
\ee
The exponent $\Psi$ can be simplified to:
\be
\Psi=-r\sum_{\alpha}\rho_{\alpha}-m\sum_{\alpha}\mu_{\alpha}
-\frac{1}{4}\sum_{\alpha\gamma}\rho_{\alpha}q^2_{\alpha\gamma}\rho_{\gamma}
+\log\bra
e^{\sum_{\gamma}\mu_{\gamma}\sigma_{\gamma}+\frac{1}{2}\sum_{\gamma\delta}q_{\gamma\delta}\rho_{\gamma}\rho_{\delta}\sigma_{\gamma}\sigma_{\delta}}\ket_{\bsigma}
\label{eq:simpleexponent}
\ee
and the distribution $D_{mr}[z]$ becomes
\be
D_{m,r}[z]=\int\frac{dx}{2\pi}e^{-\frac{1}{2}x^2+ixz}
\lim_{n\rightarrow0}
\frac{\bra
e^{-ix\sum_{\gamma}\sigma_{\gamma}\rho_{\gamma}q_{1\gamma}+\sum_{\gamma}\mu_{\gamma}\sigma_{\gamma}+\frac{1}{2}\sum_{\gamma\delta}q_{\gamma\delta}\rho_{\gamma}\rho_{\delta}\sigma_{\gamma}\sigma_{\delta}}\ket_{\bsigma}}
{\bra
e^{\sum_{\gamma}\mu_{\gamma}\sigma_{\gamma}+\frac{1}{2}\sum_{\gamma\delta}q_{\gamma\delta}\rho_{\gamma}\rho_{\delta}\sigma_{\gamma}\sigma_{\delta}}\ket_{\bsigma}}
\label{eq:RSBdist}
\ee
\vsp

The physical meaning of the order parameters $q_{\alpha\beta}$, which
in the present theory are functions of the two macroscopic state
variables $m$ and $r$,
can be
inferred in the usual manner by considering two spin systems, $\bsigma$ and
$\bsigma^{\prime}$, with the
same microscopic realisations of the quenched disorder. For such
systems we define the disorder-averaged probability distribution
$P_{mr}(q)$ for the
mutual overlap between microscopic configurations if both systems are
constrained on the same
macroscopic $(m,r)$ subshell:
\bd
P_{mr}(q)\equiv~~~~~~~~~~~~~~~~~~~~~~~~~~~~~~~~~~~~~~~~~~~~~~~~~~~~~~~~~~~~~~~~~~~~~~~~~~~~~~~~~~~~~~~~~~~~~
\ed
\bd
\bra
\frac{
\sum_{\bsigma,\bsigma^{\prime}}
\delta\left[q\minus\frac{1}{N}\sum_{k}\sigma_k\sigma^{\prime}_k\right]
\delta\left[m\minus m(\bsigma)\right]\delta\left[r\minus
r(\bsigma)\right]
\delta\left[m\minus m(\bsigma^{\prime})\right]\delta\left[r\minus
r(\bsigma^{\prime})\right]}
{
\sum_{\bsigma,\bsigma^{\prime}}
\delta\left[m\minus m(\bsigma)\right]\delta\left[r\minus
r(\bsigma)\right]
\delta\left[m\minus m(\bsigma^{\prime})\right]\delta\left[r\minus
r(\bsigma^{\prime})\right]}
\ket_{\{z_{ij}\}}
\ed
\bd
=\lim_{n\rightarrow0}\frac{1}{n(n\minus1)}\sum_{\alpha\neq\beta}
\bra\bra\delta\left[q\minus\frac{1}{N}\sum_{k}\sigma^{\alpha}_{k}\sigma^{\beta}_k\right]\prod_{\gamma=1}^n
\delta\left[m\minus m(\bsigma^{\gamma})\right]\delta\left[r\minus
r(\bsigma^{\gamma})\right]
\ket_{\{\bsigma^{\gamma}\}}\ket_{\{z_{ij}\}}
\ed
\be
=\lim_{n\rightarrow0}\frac{1}{n(n-1)}\sum_{\alpha\neq\beta}\delta\left[q-q_{\alpha\beta}\right]
\label{eq:overlapdistribution}
\ee
This dynamical equivalent
of Parisi's \cite{parisi} equilibrium order parameter function will,
in the present theory,
depend on time through the values of the two macroscopic parameters
$(m,r)$.
\vsp

The saddle-point exponent $\Psi$ that is extremised in the replica
calculation of the local field distribution has an entropic physical
interpretation. We define the
entropy per spin $\tilde{S}$ for the instantaneous macroscopic state
$(m,r)$ as
\be
\tilde{S}\equiv\lim_{N\rightarrow\infty}\frac{1}{N}\log
\sum_{\bsigma}\delta\left[m-m(\bsigma)\right]\delta\left[r-r(\bsigma)\right]
\label{eq:Stilde}
\ee
Using the replica trick $\log
Z=\lim_{n\rightarrow0}\frac{1}{n}\left[Z^n-1\right]$ and averaging over the
quenched disorder allows us to express $\tilde{S}$ in terms of the
saddle-point problem encountered in calculating $D_{mr}[z]$:
\bd
\tilde{S}=\log
2+\lim_{N\rightarrow\infty}\lim_{n\rightarrow0}\frac{1}{Nn}\left[
\bra\bra\prod_{\alpha=1}^{n}\delta\left[m-m(\bsigma^{\alpha})\right]
\delta\left[r-r(\bsigma^{\alpha})\right]\ket\ket_{\{\vec{s}^{\alpha}\},\{z_{ij}\}}-1\right]
\ed
\be
=\log 2+\lim_{n\rightarrow0}\frac{1}{n}\Psi
\label{eq:entropyisexponent}
\ee
in which $\Psi$ is the saddle-point exponent
(\ref{eq:simpleexponent}).
The entropy $\tilde{S}$ again depends on time through the values of the
macroscopic
state variables $(m,r)$.

\subsection{Equilibrium}

For large times the microscopic probability distribution
$p_t(\bsigma)$ converges to  the static Boltzmann expression
$Z^{-1}e^{-\beta H(\bsigma)}$ (with the partition function
$Z\equiv\sum_{\bsigma}e^{-\beta H(\bsigma)}$).
Since $H(\bsigma)$
(\ref{eq:energyperspin}) can be written in
terms of the macroscopic state variables $m(\bsigma)$ and
$r(\bsigma)$, at equilibrium we automatically obtain
equipartitioning of probability in the $(m,r)$ sub-shells
of the ensemble (equipartitioning in the energy shells is an
even stronger statement).
This removes the need for the second of our closure assumptions,
leaving need only for our assumption that the evolution of $m$ and $r$
is self-averaging. We will now demonstrate that  in
equilibrium we do recover the full standard results from
equilibrium statistical mechanics, including the replica symmetry
breaking (RSB) equations.

The standard replica formalism as applied to the SK model (see e.g.
\cite{mezardetal} or \cite{fisherhertz}) leads in the thermodynamic
limit $N\rightarrow\infty$ to the following
expressions for the disorder-averaged free energy per spin $\overline{f}$
\bd
\overline{f}= -\frac{1}{\beta}\log 2 + \lim_{n\rightarrow0} \min~
F(\bm,\bq)
{}~~~~~~~~~~~~~~~~~~~~~~~
\ed
\be
F(\bm,\bq)\equiv
\frac{J_0}{2n}\sum_{\alpha}m_\alpha^2
+\frac{\beta J^2}{4n}\sum_{\alpha\gamma}q^2_{\alpha\gamma}
-\frac{1}{\beta n}\log
\bra e^{\beta\sum_\alpha\sigma_\alpha(J_0 m_\alpha+\theta)
+\frac{1}{2}\beta^2
J^2\sum_{\alpha\gamma}\sigma_{\alpha}\sigma_{\gamma}q_{\alpha\gamma}}\ket_{\bsigma}
\label{eq:fequil}
\ee
The corresponding saddle point equations are
\be
m_{\gamma}=\frac{\bra \sigma_{\gamma} e^{\beta\sum_\alpha\sigma_\alpha(J_0
m_\alpha+\theta)
+\frac{1}{2}\beta^2
J^2\sum_{\alpha\gamma}\sigma_{\alpha}\sigma_{\gamma}q_{\alpha\gamma}}\ket_{\bsigma}}
{\bra e^{\beta\sum_\alpha\sigma_\alpha(J_0 m_\alpha+\theta)
+\frac{1}{2}\beta^2
J^2\sum_{\alpha\gamma}\sigma_{\alpha}\sigma_{\gamma}q_{\alpha\gamma}}\ket_{\bsigma}}
\label{eq:saddlemequil}
\ee
\be
q_{\gamma\delta}=\frac{\bra \sigma_{\gamma}\sigma_{\delta}
e^{\beta\sum_\alpha\sigma_\alpha(J_0 m_\alpha+\theta)
+\frac{1}{2}\beta^2
J^2\sum_{\alpha\gamma}\sigma_{\alpha}\sigma_{\gamma}q_{\alpha\gamma}}\ket_{\bsigma}}
{\bra e^{\beta\sum_\alpha\sigma_\alpha(J_0 m_\alpha+\theta)
+\frac{1}{2}\beta^2
J^2\sum_{\alpha\gamma}\sigma_{\alpha}\sigma_{\gamma}q_{\alpha\gamma}}\ket_{\bsigma}}
\label{eq:saddleqequil}
\ee
and the physical interpretation in terms of the two
(disorder-averaged) functions $P(q)$ and $P(m)$ is:
\bd
P(q)\equiv
\bra
Z^{-2}
\sum_{\bsigma\bsigma^{\prime}}
\delta\left[q\minus\frac{1}{N}\sum_{k}\sigma_k\sigma^{\prime}_k\right]
e^{-\beta H(\bsigma)-\beta H(\bsigma^{\prime})}
\ket_{\{z_{ij}\}}
\ed
\be
=\lim_{n\rightarrow0}\frac{1}{n(n-1)}\sum_{\alpha\neq\gamma}\delta\left[q-q_{\alpha\gamma}\right]
\label{eq:overlapdistrequil}
\ee
\bd
P(m)\equiv
\bra
Z^{-1}
\sum_{\bsigma}
\delta\left[m\minus\frac{1}{N}\sum_{k}\sigma_k\right]
e^{-\beta H(\bsigma)}
\ket_{\{z_{ij}\}}
\ed
\be
=\lim_{n\rightarrow0}\frac{1}{n}\sum_{\alpha}\delta\left[m-m_{\alpha}\right]
\label{eq:magndistrequil}
\ee
According to Parisi's \cite{parisi} theory the magnetisation is
self-averaging, even in the regime where replica symmetry is broken
\cite{mezardvirasoro},
so $P(m)$ is a delta-function and $m_{\alpha}=m$ for all $\alpha$.
{}From the internal energy
in thermal equilibrium $E/N=[1+\beta\partial_{\beta}]\overline{f}$, which is
also self-averaging
\cite{mezardvirasoro}, we obtain the equilibrium
expression for our dynamic order parameter $r$:
\bd
r_{\rm eq}=\frac{1}{2}\beta J\left[1-\int\!dq~P(q)q^2\right]
\ed
For $\beta J_0<1$, where $m_{\rm eq}=0$, the continuous transition at
$\beta J=1$ from the
paramagnetic phase with $P(q)=\delta(q)$ to the spin-glass phase,
 is therefore marked by
$r_{\rm eq}=\frac{1}{2}$.

Comparison with the dynamical eqns.
(\ref{eq:saddle_m},\ref{eq:saddle_q},\ref{eq:saddle_r}), shows that
the two approaches
yield identical equations  if we impose the following conditions:
\be
\mu_{\alpha}=\mu\equiv\beta(J_0 m+\theta)~~~~~~~~\rho_\alpha=\rho\equiv\beta
J
\label{eq:conditions}
\ee
Below we show that these conditions turn out to be precisely those
which imply
dynamical stability with respect to the macroscopic flow
(\ref{eq:mequation},\ref{eq:requation});
\bd
\frac{d}{dt}m=0,~~~~~~\frac{d}{dt}r=0
\ed
and hence they also describe the same equilibrium physics.

First we consider the evolution of $m$, using the noise distribution
(\ref{eq:RSBdist}) and the conditions (\ref{eq:conditions}). If we
perform a shift of the integration line for $z$ and perform the
integral over $x$ we arrive at:
\bd
\frac{d}{dt}m=-m+\lim_{n\rightarrow0}\int\!Dz~
\frac
{\bra\tanh[\rho z\plus\mu\plus\rho^2\sum_{\alpha}q_{1\alpha}\sigma_{\alpha}]~
e^{\mu\sum_{\gamma}\sigma_{\gamma}+\frac{1}{2}\rho^2\sum_{\gamma\delta}q_{\gamma\delta}\sigma_{\gamma}\sigma_{\delta}}\ket_{\bsigma}}
{\bra
e^{\mu\sum_{\gamma}\sigma_{\gamma}+\frac{1}{2}\rho^2\sum_{\gamma\delta}q_{\gamma\delta}\sigma_{\gamma}\sigma_{\delta}}\ket_{\bsigma}}
\ed
with the abbreviation
$Dz\equiv(2\pi)^{-\frac{1}{2}}e^{-\frac{1}{2}z^2}dz$. In the
numerator of this
expression we perform the average over $\sigma_1$ explicitly, and use the
identity
\be
e^{-u}\int\!Dz~\tanh[\rho z\minus\rho^2\plus u]+
e^{u}\int\!Dz~\tanh[\rho z\plus\rho^2\plus u]=2\sinh[u]
\label{eq:identity1}
\ee
to arrive at:
\bd
\frac{d}{dt}m=-m+\lim_{n\rightarrow0}
\frac
{\bra\sigma_1~
e^{\mu\sum_{\gamma}\sigma_{\gamma}+\frac{1}{2}\rho^2\sum_{\gamma\delta}q_{\gamma\delta}\sigma_{\gamma}\sigma_{\delta}}\ket_{\bsigma}}
{\bra
e^{\mu\sum_{\gamma}\sigma_{\gamma}+\frac{1}{2}\rho^2\sum_{\gamma\delta}q_{\gamma\delta}\sigma_{\gamma}\sigma_{\delta}}\ket_{\bsigma}}=0
\ed
(utilizing (\ref{eq:saddle_m})).

In a similar way we obtain for the evolution of $r$:
\bd
\frac{d}{dt}r=-2r+\lim_{n\rightarrow0}\int\!Dz~~~~~~~~~~~~~~~~~~~~~~~~~~~~~~~~~~~~~~~~~~~~~~~~~~~~~~~~~~~~~~~~~~~~~~~~~~~~~~~~
\ed
\bd
\frac
{\bra[z\plus\rho\sum_{\alpha}q_{1\alpha}\sigma_{\alpha}]\tanh[\rho
z\plus\mu\plus\rho^2\sum_{\alpha}q_{1\alpha}\sigma_{\alpha}]~
e^{\mu\sum_{\gamma}\sigma_{\gamma}+\frac{1}{2}\rho^2\sum_{\gamma\delta}q_{\gamma\delta}\sigma_{\gamma}\sigma_{\delta}}\ket_{\bsigma}}
{\bra
e^{\mu\sum_{\gamma}\sigma_{\gamma}+\frac{1}{2}\rho^2\sum_{\gamma\delta}q_{\gamma\delta}\sigma_{\gamma}\sigma_{\delta}}\ket_{\bsigma}}
\ed
Again we perform the average over $\sigma_1$ in the numerator
explicitly, simplify the result with the identity
\bd
e^{-u}\int\!Dz~[\rho z\minus\rho^2\plus u]\tanh[\rho z\minus\rho^2\plus u]+
e^{u}\int\!Dz~[\rho z\plus\rho^2\plus u]\tanh[\rho z\plus\rho^2\plus
u]
\ed
\be
=2u\sinh[u]+_2\rho^2\cosh[u]
\label{eq:identity2}
\ee
and arrive at:
\bd
\frac{d}{dt}r=-2r+\lim_{n\rightarrow0}\left\{
\rho+\rho\sum_{\alpha>1}q_{1\alpha}
\frac
{\bra\sigma_1\sigma_{\alpha}~
e^{\mu\sum_{\gamma}\sigma_{\gamma}+\frac{1}{2}\rho^2\sum_{\gamma\delta}q_{\gamma\delta}\sigma_{\gamma}\sigma_{\delta}}\ket_{\bsigma}}
{\bra
e^{\mu\sum_{\gamma}\sigma_{\gamma}+\frac{1}{2}\rho^2\sum_{\gamma\delta}q_{\gamma\delta}\sigma_{\gamma}\sigma_{\delta}}\ket_{\bsigma}}\right\}
\ed
\bd
=\lim_{n\rightarrow0}\rho\sum_{\alpha}q_{1\alpha}^2-2r=0
\ed
(utilizing (\ref{eq:saddle_q},\ref{eq:saddle_r})).

Finally we use the equilibrium conditions (\ref{eq:conditions}) to show that
the thermodynamic entropy per spin
$S=\beta^2\partial_{\beta}\overline{f}$ in equilibrium coincides with
the dynamic entropy per spin $\tilde{S}$ given by
(\ref{eq:Stilde}):
\bd
S=
\log 2 - \lim_{n\rightarrow0}\left\{
m\mu+\frac{3\rho^2}{4n}\sum_{\alpha\gamma}q_{\alpha\gamma}^2
-\frac{1}{n}\log
\bra e^{\mu\sum_\alpha\sigma_\alpha
+\frac{1}{2}\rho^2\sum_{\alpha\gamma}\sigma_{\alpha}\sigma_{\gamma}q_{\alpha\gamma}}\ket_{\bsigma}\right\}
\ed
According to (\ref{eq:simpleexponent}) and
(\ref{eq:entropyisexponent}) this expression is identical to the
one we obtained for $\tilde{S}$.

\section{Replica Symmetry}

\subsection{Replica-Symmetric Local Field Distribution}

We first make the replica-symmetric ansatz (RS) and assume
$P_{mr}(q)$ (\ref{eq:overlapdistribution}) to be a delta-function, so
$q_{\alpha\beta}=\delta_{\alpha\beta}+q(1-\delta_{\alpha\beta})$. From
this ansatz the saddle-point equations
(\ref{eq:saddle_m},\ref{eq:saddle_q},\ref{eq:saddle_r}) allow us to
deduce $\mu_{\alpha}=\mu$ and $\rho_{\alpha}=\rho$. For
$n\rightarrow0$ we obtain:
\be
m=\int\!Du~\tanh(\rho\sqrt{q}u+\mu)
\label{eq:RSsaddle_m}
\ee
\be
q=\int\!Du~\tanh^2(\rho\sqrt{q}u+\mu)
\label{eq:RSsaddle_q}
\ee
\be
\rho=\frac{2r}{1-q^2}
\label{eq:RSsaddle_r}
\ee
The corresponding local field
distribution $D^{RS}_{mr}[z]$ becomes:
\bd
D^{RS}_{mr}[z]=\int\frac{dx}{2\pi}e^{-\frac{1}{2}x^2+ixz}
\lim_{n\rightarrow0}
\int\!Du~\cosh\left[\rho\sqrt{q}u\plus\mu\minus ix\rho\right]
\cosh^{n-1}\left[\rho\sqrt{q}u\plus\mu\minus ixq\rho\right]
\ed
We first perform the shift
$u\rightarrow v+ix\sqrt{q}$, after which the limit $n\rightarrow0$ can be
safely taken. In the resulting expression we can perform the integral over $x$.
After some final
transformations of integration variables we arrive at
\bd
D_{mr}^{RS}[z]=\frac{e^{-\frac{1}{2}\left[z+\rho(1-q)\right]^2}}{2\sqrt{2\pi}}
\left\{1+\int\!Dy~\tanh\left[\rho y\sqrt{q(1\minus q)}\minus\rho
q\left[z\plus\rho(1\minus q)\right]\minus\mu\right]\right\}
\ed
\be
+\frac{e^{-\frac{1}{2}\left[z-\rho(1-q)\right]^2}}{2\sqrt{2\pi}}
\left\{1+\int\!Dy~\tanh\left[\rho y\sqrt{q(1\minus q)}\plus\rho
q\left[z\minus\rho(1\minus q)\right]\plus\mu\right]\right\}
\label{eq:finalRSdistribution}
\ee
This expression cannot be simplified further, except for three special
cases which we will discuss below.\vsp

{}From expression (\ref{eq:finalRSdistribution}) and the saddle-point
equations it is clear that
$D_{mr}^{RS}[z]$ is Gaussian only along the line $r=0$:
\be
r=0:~~~~~~~~~~
D_{m,0}^{RS}[z]=\frac{1}{\sqrt{2\pi}}e^{-\frac{1}{2}z^2}
\label{eq:RSdistr0}
\ee
For $r=0$ we obtain $q=m^2$.
We can identify such macroscopic
states as purely ferromagnetic (for $m\neq0$) or paramagnetic (for $m=0$).
The  result (\ref{eq:RSdistr0}) is indeed what one would obtain in
thermal equilibrium for
$\beta J=0$ (where only the para-magnetic and purely ferromagnetic
states are found).

A second simplification of (\ref{eq:finalRSdistribution}) results
for $q=0$ (the paramagnetic state), which can only occur along the line
$m=0$.
For $m=0$ the RS saddle-point equations reduce to
\bd
q=F(q)\equiv\int\!Du~\tanh^2\left[\frac{2ur\sqrt{q}}{1-q^2}\right]
\ed
with the properties
\bd
F(1)=1~~~~~~F(q)=4r^2q-32r^4q^2+\order(q^3)
\ed
from which we conclude that along the $m=0$ line we find a
paramagnetic ($q=0$) state for $r<\frac{1}{2}$:
\be
m=0,~r<\frac{1}{2}:~~~~~~~~~~
D_{0,r}^{RS}[z]=
\frac{1}{2\sqrt{2\pi}}e^{-\frac{1}{2}[z+2r]^2}+
\frac{1}{2\sqrt{2\pi}}e^{-\frac{1}{2}[z-2r]^2}
\label{eq:RSdistrm0}
\ee
This result is
indeed what one would obtain for the field distribution in thermal equilibrium
in the
paramagnetic region of the phase diagram \cite{thomsenetal}. For
$r>\frac{1}{2}$, $m=0$ we obtain a spin-glass with $q\neq 0$, where
again we know from equilibrium studies \cite{thomsenetal} that the
local field distribution indeed has a non-trivial form like that in
(\ref{eq:finalRSdistribution}).

The third simplification occurs for $q\approx1$. Expanding the
saddle-point equations in
powers of $\epsilon\equiv1-q$ gives the leading orders
\be
\rho=r\epsilon^{-1}+\ldots~~~~~~\mu=r\epsilon^{-1}\sqrt{2}~\erf^{-1}(m)+\ldots
\label{eq:scalingnearq1}
\ee
\be
r=\sqrt{\frac{2}{\pi}}e^{-\left[\erf^{-1}(m)\right]^2}
\label{eq:q1line}
\ee
Equation (\ref{eq:q1line}) defines the line in the $(m,r)$ plane where the
situation $q=1$ actually occurs. Near this line we can use the scaling
relations (\ref{eq:scalingnearq1}) to show that expression
(\ref{eq:finalRSdistribution}) reduces to the Schowalter-Klein
\cite{schowalterklein} form, which in equilibrium would be obtained
in the limit of zero temperature \cite{thomsenetal} (in RS approximation):
\bd
D_{m,r(m)}^{RS}[z]
=
\frac{e^{-\frac{1}{2}[z+r(m)]^2}}{\sqrt{2\pi}}
\theta\left[\minus z\minus r(m)\minus\sqrt{2}~\erf^{-1}(m)\right]~~~~~~~~~~
\ed
\be
{}~~~~~~~~~~+\frac{e^{-\frac{1}{2}[z- r(m)]^2}}{\sqrt{2\pi}}
\theta\left[z\minus r(m)\plus\sqrt{2}~\erf^{-1}(m)\right]
\label{eq:Rsdistrq1}
\ee
in which $r(m)$ denotes the $q=1$ line (\ref{eq:q1line}).

\subsection{Special Lines in the Flow Diagram}

In order to check the applicability of the RS ansatz we
calculate the equivalent of the RS zero-entropy (`freezing') line in the
$(m,r)$ plane (where the number of microscopic
configurations contributing to our averages vanishes), and the de
Almeida-Thouless (AT)
line \cite{AT}, where a replica-symmetry breaking (RSB) solution of the
saddle-point equations bifurcates from the RS saddle-point.

In RS theory the dynamic entropy
(\ref{eq:Stilde}) is, according to (\ref{eq:entropyisexponent}), given by
\be
\tilde{S}_{RS}=\log 2+\int\!Du~\log\cosh\left[\rho
u\sqrt{q}\plus\mu\right]-m\mu+\frac{1}{4}\rho^2(1\minus q)^2-\rho r
\label{eq:RSStilde}
\ee
For
$r=0$ (where there is no spin-glass alignment) the entropy reduces to
\bd
\tilde{S}_{RS,~r=0}=\log 2-\int_0^{\tanh^{-1}(|m|)}ds
{}~s\left[1-\tanh^2(s)\right]~\in~[0,\log2]
\ed
with $\tilde{S}_{RS,~r=0}=\log2$ at $m=0$ (the para-magnetic state)
down to $\tilde{S}_{RS,~r=0}=0$ at $m=\pm 1$ (the fully ordered
ferro-magnetic state). Along the line $m=0$, below $r=\frac{1}{2}$, we
find $q=0$ and $\tilde{S}_{RS}=\log2-r^2>0$. Using the scaling
relations (\ref{eq:scalingnearq1}) one can finally show that near the
$q=1$ line (\ref{eq:q1line}) the RS entropy is negative,
except for $|m|=1,~r=0$, where the $q=1$ line and the line
$\tilde{S}_{RS}=0$ meet. Since the physical dynamical entropy cannot
be negative this already signals an inadequacy in the RS ansatz,
analogous to that found in the equilibrium RS theory of SK \cite{SK}.
The full curve $\tilde{S}_{RS}=0$ signals this inadequacy in $(m,r)$ space.
\begin{figure}
\vspace*{25cm}
\hbox to \hsize{\hspace*{-40mm}\includegraphics{ps/lines.ps}\hspace*{1cm}}
\vspace*{-15cm}
\caption{AT line (large dashes), RS freezing line (dashes/dots) and
$q=1$ line (small dashes) in the $(m,r)$ plane.}
\label{fig:lines}
\end{figure}

An AT-line \cite{AT} signals the first continuous bifurcation of a saddle-point
solution
without replica symmetry from the replica-symmetric one. We follow the
usual convention and assume that the first such bifurcation is the
replicon mode:
\bd
q_{\alpha\beta}\rightarrow q+\delta
q_{\alpha\beta},~~~~\rho_{\alpha}=\rho,~~~~\mu_{\alpha}=\mu
\ed
Inserting this ansatz into the full saddle point equations
shows that the RSB bifurcations are of the form $\sum_{\alpha\neq\beta}\delta
q_{\alpha\beta}=0$. After some
bookkeeping and after taking the limit $n\rightarrow0$ one then obtains the
bifurcation condition which
defines the dynamic AT line:
\be
1-\rho^2\int\! Du~\cosh^{-4}[\rho\sqrt{q}y+\mu]=0
\label{eq:ATline}
\ee
The RS solution
is stable as long as the left-hand side of (\ref{eq:ATline}) is
positive. For $r=0$ (with $|m|<1$) the RS
solution is indeed stable. The AT line intersects the line $m=0$ at
$r=\frac{1}{2}$.
Using the scaling
relations (\ref{eq:scalingnearq1}) one can also show that near the
$q=1$ line (\ref{eq:q1line}) the RS solution is unstable,
except for $|m|=1,~r=0$, where the $q=1$ line and the AT line meet.

In figure \ref{fig:lines} we show the freezing line (where
$\tilde{S}_{RS}=0$) (\ref{eq:RSStilde}), the AT line (\ref{eq:ATline}) and
the $q=1$ line (\ref{eq:q1line}) in the $(m,r)$ plane.
We note that the $q=1$ line always lies above the $\tilde{S}_{RS}=0$
line, which in turn lies above the AT line, except at $|m|=1$, $r=0$.
Thus the AT line is the critical one for replica symmetry. The
separation between the AT line and the $q=1$ line, which provides
an effective boundary for the $(m,r)$ dynamics, is greatest for small
$m$ where the ferromagnetic order is small and occurs for large $r$,
when spin-glass alignment is greatest.

Below the AT line the RS solution is stable against RSB fluctuations.
The RS solution breaks down in the region where ferromagnetic order is
small  and spin-glass type field-alignment dominates.

\subsection{Replica-Symmetric Flow Equations}

By combining the equations (\ref{eq:mequation},\ref{eq:requation}) with
expression (\ref{eq:finalRSdistribution}) we arrive at a closed set of
autonomous differential
equations describing the deterministic evolution of the macroscopic
state $(m,r)$:
\be
\frac{d}{dt}m=\int\!\!\int\!DxDy~M(m,r;x,y)-m
\label{eq:finaldv_m}
\ee
\be
\frac{d}{dt}r=\int\!\!\int\!DxDy~R(m,r;x,y)-2r
\label{eq:finaldv_r}
\ee
in which
\bd
M(m,r;x,y)=~~~~~~~~~~~~~~~~~~~~~~~~~~~~~~~~~~~~~~~~~~~~~~~~~~~~~~~~~~~~~
{}~~~~~~~~~~~~~~~~~~~~~~~~~~~~~~~~~~~~~~
\ed
\bd
\frac{1}{2}
\left[1\minus\tanh\left[x\rho\sqrt{q(1\minus q)}\plus\rho q
y\plus\mu\right]\right]\tanh\beta\left[J_0m\plus Jy\plus\theta\minus
J\rho(1\minus q)\right]
\ed
\bd
+\frac{1}{2}
\left[1\plus\tanh\left[x\rho\sqrt{q(1\minus q)}\plus\rho q
y\plus\mu\right]\right]\tanh\beta\left[J_0m\plus Jy\plus\theta\plus
J\rho(1\minus q)\right]
\ed
\bd
R(m,r;x,y)=~~~~~~~~~~~~~~~~~~~~~~~~~~~~~~~~~~~~~~~~~~~~~~~~~~~~~~~~~~~
{}~~~~~~~~~~~~~~~~~~~~~~~~~~~~~~~~~~~~~~~
\ed
\bd
\frac{1}{2}\left[y\minus\rho(1\minus q)\right]
\left[1\minus\tanh\left[x\rho\sqrt{q(1\minus q)}\plus\rho q
y\plus\mu\right]\right]\tanh\beta\left[J_0m\plus Jy\plus\theta\minus
J\rho(1\minus q)\right]
\ed
\bd
+\frac{1}{2}\left[y\plus\rho(1\minus q)\right]
\left[1\plus\tanh\left[x\rho\sqrt{q(1\minus q)}\plus\rho q
y\plus\mu\right]\right]\tanh\beta\left[J_0m\plus Jy\plus\theta\plus
J\rho(1\minus q)\right]
\ed
with $\{q,\rho,\mu\}$ being functions of the macroscopic
state $(m,r)$, to be solved from the saddle-point equations
(\ref{eq:RSsaddle_m},\ref{eq:RSsaddle_q},\ref{eq:RSsaddle_r}).

\begin{figure}
\vspace*{24cm}
\hbox to \hsize{\hspace*{-4cm}\includegraphics{ps/flow.ps}\hspace*{6cm}}
\label{fig:flow}
\end{figure}
\begin{figure}
\caption{Trajectories in the $(m,r)$ plane obtained by
performing sequential simulations of the SK model with $N=3000$ and
zero external field ,
for $t\leq10$ iterations/spin (solid lines),
together with the velocities
as predicted by the theory (arrows, calculated at intervals of 1
iteration/spin for the instantaneous macroscopic state of the
corresponding simulation, at the point of the base of the arrow). The first row
of graphs corresponds to
$T=1.5$, the second to $T=1.0$, the third to $T=0.5$ and the fourth to
$T=0$. Dashed lines indicate the $q=1$ line (upper), the RS freezing
line (middle) and
the AT line (lower).}
\end{figure}

In figure \ref{fig:flow} we compare the flow defined by
(\ref{eq:finaldv_m},\ref{eq:finaldv_r}) with numerical simulations for
$N=3000$, $\theta=0$, $J=1$, $J_0\in\{0,1,2\}$ and four choices of the
temperature
$T$.
The parameters $J_0$ and $T$ have been chosen in such a way that the
corresponding equilibrium situations (according to standard equilibrium theory
\cite{fisherhertz}) include spin-glass states ($J_0<1$, $T<1$),
states with ferro-magnetic order ($J_0>1$, $T<J_0$)
and para-magnetic states ($J_0<T$, $T>1$).
 At
intervals of $\Delta t=1$ iteration/spin we measure the
macroscopic order parameters $(m,r)$ in the simulated system and
calculate the derivatives
$(\frac{d}{dt}m,\frac{d}{dt}r)$ as predicted by
(\ref{eq:finaldv_m},\ref{eq:finaldv_r}). The initial states generating
the
trajectories (labelled by $\ell=0,\ldots,10$) were drawn
at random according to
$p_0(\vec{s})\equiv\prod_i\left[\frac{1}{2}[1\plus\frac{\ell}{10}]\delta_{s_i,\xi_{i}^1}\plus\frac{1}{2}[1\minus\frac{\ell}{10}]\delta_{s_i,\minus\xi_{i}^{1}}\right]$,
such that
that $\bra m\ket_{t=0}=0.1\ell$ and $\bra
r\ket_{t=0}=1$.
The figure indicates that the flow is described quite well by
(\ref{eq:finaldv_m},\ref{eq:finaldv_r}), except for those regions in
the $(m,r)$ plane where the RS solution is unstable (above the
AT line). More detailed comparisons between theory and simulations
will be made in a subsequent section.
\vsp

{}From the RS saddle-point equations
(\ref{eq:RSsaddle_m},\ref{eq:RSsaddle_q},\ref{eq:RSsaddle_r}) we can
directly recover all equilibrium results obtained by Sherrington and
Kirkpatrick \cite{SK,KS}.
Inserting
the two relations $\rho=\beta J$ and $\mu=\beta(J_0 m+\theta)$ into our RS
saddle-point equations gives
\bd
m=\int\!Du~\tanh\beta(J_0 m\plus J\sqrt{q}u\plus\theta)
\ed
\bd
q=\int\!Du~\tanh^2\beta(J_0 m\plus J\sqrt{q}u\plus\theta)
\ed
\bd
r=\frac{1}{2}\beta J\left[1-q^2\right]
\ed
We now use the identities (\ref{eq:identity1},\ref{eq:identity2})
and perform a rotation in the space of the gaussian integrals in
(\ref{eq:finaldv_m},\ref{eq:finaldv_r}) to arrive for the RS thermal
equilibrium state of \cite{SK,KS} at
\bd
\frac{d}{dt}m=\int\!Dx~\tanh\beta(J_0 m\plus J\sqrt{q}x\plus\theta)-m=0
\ed
\bd
\frac{d}{dt}r=\beta Jq\left[q- \int\!Dx~\tanh^2\beta(J_0 m\plus
J\sqrt{q}x\plus\theta)\right]=0
\ed
The RS order parameter equations in thermal equilibrium, as derived in
\cite{SK,KS}, thus indeed define fixed-points of our flow equations,
as also follows from our more general analysis of section 2.4.

If we insert the fixed-point relations into our expression
(\ref{eq:ATline}) for the AT line, we obtain:
\bd
1
=\beta^2 J^2\int\!Dx~\cosh^{-4}\beta(J_0 m\plus J\sqrt{q}x\plus\theta)
\ed
which, again, corresponds exactly to the result obtained in thermal
equilibrium \cite{AT}.  This includes both the line segment separating
paramagnetic
from spin-glass phase, where $(m_{\rm eq},r_{\rm eq})=(0,\frac{1}{2})$,
and the line segment separating the replica-symmetric and
replica-symmetry broken ferromagnetic phases.

Our dynamical RS laws (\ref{eq:finaldv_m},\ref{eq:finaldv_r}) thus lead
precisely to the thermal equilibrium described by Sherrington and
Kirkpatrick \cite{SK,KS} and de Almeida and Thouless \cite{AT}, including
entropy and stability
with respect to replica-symmetry breaking.

\subsection{Relaxation Times}

We will investigate the asymptotic behaviour of the RS flow equations.
For simplicity and to suppress notation
we restrict ourselves to the case $J_0=0$, $J=1$.
We expand both RS flow
equations (\ref{eq:mequation},\ref{eq:requation}) around the
equilibrium state $(m,r)=(0,\req)$
\bd
m(t)=\epsilon\tilde{m}(t)+\order(\epsilon^2)~~~~~~r(t)=\req-\epsilon\tilde{r}(t)+\order(\epsilon^2)
\ed
\be
\req=\frac{1}{2}\beta(1-q^2)~~~~~~~~q=\int\!Dy~\tanh^2\left[\beta\sqrt{q}y\right]
\label{eq:m0equil}
\ee
as well as the RS
saddle-point equations
(\ref{eq:RSsaddle_m},\ref{eq:RSsaddle_q},\ref{eq:RSsaddle_r}):
\be
\left(\frac{\partial q}{\partial m}\right)_{0,\req}\!\!=0~~~~~~
\left(\frac{\partial\mu}{\partial m}\right)_{0,\req}\!\!=\frac{1}{1-q}~~~~~~
\left(\frac{\partial
\mu}{\partial r}\right)_{0,\req}\!\!=0
\label{eq:saddlederivs1}
\ee
\be
\left(\frac{\partial q}{\partial r}\right)_{0,\req}\!\!=
\frac{4q}{\beta(1\plus 3q^2)}~\frac
{\int\!Du~u^2\tanh^2(\beta u\sqrt{q})-q}
{2q(1-q^2)(1+3q^2)^{-2}-\int\!Du~ u^2\tanh^2(\beta u\sqrt{q})+q}
\label{eq:saddlederivs2}
\ee
The linearised flow equations
decouple since
$\left(\partial_m D\right)^{RS}_{0,r}[z]$ and
$\left(\partial_r D\right)^{RS}_{0,r}[z]$ are respectively  anti-symmetric and
symmetric in $z$ (this decoupling does not
depend on replica symmetry).
As a result we directly obtain the two relaxation times:
\be
\tau_m^{-1}=
-\lim_{t\rightarrow\infty}\frac{1}{t}\log\left[\frac{\tilde{m}(t)}{\tilde{m}(0)}\right]=
1-\int\!dz \left(\partial_m
D\right)^{RS}_{0,\req}[z]\tanh\left[\beta z\right]
\label{eq:tau_m}
\ee
\be
\tau_r^{-1}=
-\lim_{t\rightarrow\infty}\frac{1}{t}\log\left[\frac{\tilde{r}(t)}{\tilde{r}(0)}\right]=
2- \int\!dz \left(\partial_r D\right)^{RS}_{0,\req}[z]z\tanh\left[\beta
z\right]
\label{eq:tau_r}
\ee
In the paramagnetic temperature
region $T>1$ the partial derivatives of the local field
distribution (\ref{eq:finalRSdistribution}) can be calculated easily.
We find
\be
\tau_m^{-1}=1-\int\!Dz~\tanh(\beta z\plus
\beta^2)
\label{eq:paratau_m}
\ee
\be
\tau_r^{-1}=2-2\int\!Dz~z
(z\plus\beta)\tanh(\beta z\plus
\beta^2)
\label{eq:paratau_r}
\ee
For $T<1$ more work is required to find the partial derivatives
$(\partial D)$ and the relaxation times.
\vsp

\begin{figure}
\vspace*{11cm}
\hbox to \hsize{\hspace*{-45mm}\includegraphics{ps/relaxtimes.ps}\hspace*{45mm}}
\vspace*{-3cm}
\caption{The two asymptotic RS relaxation times $\tau_m$ and $\tau_r$
for $J_0=\theta=0$ and $J=1$ as a function of
temperature.}
\label{fig:relaxtimes}
\end{figure}
We first turn to the magnetisation.
After some bookkeeping we can derive from
 (\ref{eq:finalRSdistribution}):
\bd
\left(\partial_m
D\right)_{0,\req}[z]=
\frac{e^{-\frac{1}{2}[z-\Delta]^2}}{2(1\minus
q)\sqrt{2\pi}}\left\{1-\int\!Dy\tanh^2\Delta\left[
y\left(\frac{q}{1\minus q}\right)^{\frac{1}{2}}
\plus(z\minus\Delta)\frac{q}{1\minus q}\right]\right\}
\ed
\bd
-\frac{e^{-\frac{1}{2}[z+\Delta]^2}}{2(1\minus
q)\sqrt{2\pi}}\left\{1-\int\!Dy\tanh^2\Delta\left[
y\left(\frac{q}{1\minus
q}\right)^{\frac{1}{2}}\plus(z\plus\Delta)\frac{q}{1\minus
q}\right]\right\}
\ed
where $\Delta\equiv\beta(1-q)$. With this expression we obtain, after a
rotation in the space of the integrals:
\be
\tau_m=\frac{1-q}
{\int\!Dy\left[1\minus\tanh^2(\beta y\sqrt{q})\right]\int\! Dz\left[1\minus
\tanh\beta\left(y\sqrt{q}\plus z\sqrt{1-q}\plus\beta(1\minus q)\right)\right]}
\label{eq:SGtau_m}
\ee
For non-zero temperatures the asymptotic relaxation of $m$
described by the RS equations is indeed exponential.
For $T\rightarrow0$,
however, we can use $q\sim1-\beta^{-1}\sqrt\frac{2}{\pi}$ to show that the
relaxation becomes non-exponential:
\bd
\lim_{\beta\rightarrow\infty}\tau_m^{-1}=
1-\frac{1}{2}\int\!dz~e^{-\frac{1}{2}z^2}\sgn[z\plus\sqrt\frac{2}{\pi}]~\frac{d}{dz}\sgn[z]=0
\ed
Next we turn to the relaxation of
$r$. Taking the appropriate derivatives results in
\bd
\tau_r^{-1}=2
-\frac{2}{1\plus q}
\int\!Dy\left[1\plus\tanh(\beta y\sqrt{q})\right]
\int\!Dz\left[\tanh(\beta Q)\plus\beta
Q\left[1\minus\tanh^2(\beta Q)\right]\right]
\ed
\bd
-\frac{2\sqrt{q}}{1\minus
q^2}\left[1\plus\frac{\beta}{4q}(1\plus3q^2)
\left(\frac{\partial q}{\partial r}\right)_{\rm eq}\right]
\int\!Dy~y\left[1\minus\tanh^2(\beta y\sqrt{q})\right]
\int\!Dz~Q\tanh(\beta Q)
\ed
\bd
-\left(\frac{\partial q}{\partial r}\right)_{\rm eq}
\int\!Dy\left[1\plus\tanh(\beta y\sqrt{q})\right]
\int\!Dz\left[\frac{y}{2\sqrt{q}}\minus\frac{z}{2\sqrt{1\minus
q}}\minus\frac{\beta(1\minus q^2)}{(1\plus q)^2}\right]
\ed
\be
\times
\left[\tanh(\beta Q)\plus\beta
Q\left[1\minus\tanh^2(\beta Q)\right]\right]
\label{eq:SGtau_r}
\ee
with the abbreviation $Q\equiv y\sqrt(q)\plus z\sqrt{1\minus q}\plus
\beta(1\minus q)$ (to be used in combination with (\ref{eq:saddlederivs2})).
Numerical evaulation of the integrals in
(\ref{eq:paratau_m},\ref{eq:paratau_r},\ref{eq:SGtau_m},\ref{eq:SGtau_r}) as a
function of temperature results in figure
\ref{fig:relaxtimes}. Note that in the absence of spin-glass
interactions (i.e. for $J=0$) one would simply find
$\tau_m=1$ for all $T$. Both relaxation times diverge for
$T\rightarrow0$, with
$\lim_{T\rightarrow0}\tau_r/\tau_m=0$.

\section{Further Comparisons with Numerical Simulations}

In this section we present some more detailed simulation experiments,
the outcome of which is compared to the predictions of our RS theory
(the latter need not give sensible results above the AT line). The simulations
can display two
types of finite size effects: thermal fluctuations in the flow of
the order parameters (i.e. finite size
corrections
to the Liouville equation (\ref{eq:secondmacromaster})) and
fluctuations in the local field distribution
(i.e. finite size corrections to the steepest descent integration
leading to (\ref{eq:secondmacromaster})).

\subsection{The Local Field Distribution}

\begin{figure}
\vspace*{25cm}
\hbox to \hsize{\hspace*{-40mm}\includegraphics{ps/distflow.ps}\hspace*{1cm}}
\vspace*{-15cm}
\caption{Trajectories in the $(m,r)$ plane obtained from
 sequential simulations of the SK model with $N=3200$,
$J=1$ and $T=0.1$, for three different choices of $J_0$. Initial states:
$(m,r)\sim(0.5,0)$. Dots indicate times at which the spin-glass
contributions to the local fields are measured in order to test the
theory.}
\label{fig:distflow}
\end{figure}
\begin{figure}
\vspace*{21cm}
\hbox to \hsize{\hspace*{-45mm}\includegraphics{ps/comparefer0.ps}\hspace*{1cm}}
\vspace*{-4cm}
\caption{Comparison between RS theory (dashed lines) and the local
field
distribution as measured during the $J_0=0$ simulation.}
\label{fig:compare0}
\end{figure}
\begin{figure}
\vspace*{21cm}
\hbox to \hsize{\hspace*{-45mm}\includegraphics{ps/comparefer1.ps}\hspace*{1cm}}
\vspace*{-4cm}
\caption{Comparison between RS theory (dashed lines) and the local
field
distribution as measured during the $J_0=1$ simulation.}
\label{fig:compare1}
\end{figure}
\begin{figure}
\vspace*{21cm}
\hbox to \hsize{\hspace*{-45mm}\includegraphics{ps/comparefer2.ps}\hspace*{1cm}}
\vspace*{-4cm}
\caption{Comparison between RS theory (dashed lines) and the local
field
distribution as measured during the $J_0=2$ simulation.}
\label{fig:compare2}
\end{figure}

First we compare our analytical result (\ref{eq:finalRSdistribution})
directly with
the outcome of measuring the spin-glass contributions to the local
alignment fields during
actual numerical simulations. In order to probe the different regions
of the $(m,r)$ plane we performed simulations from the initial state
$(m,r)\sim(0.5,0)$ for $J_0=0$, $J_0=1$ and $J_0=2$ and measured the
instantaneous distribution of the spin-glass contributions to the
local alignment fields at different times. In figure
\ref{fig:distflow} we show the resulting trajectories in the
$(m,r)$ plane (solid lines), together with the AT line (lower dashed
line) and the $q=1$ line (upper dashed line). Dots indicate the instances were
the relevant measurements
were done: $t=0$, $t=1$, $t=5$ and $t=10$ (unit: iterations per spin).
In figures \ref{fig:compare0}, \ref{fig:compare1} and
\ref{fig:compare2} the distributions as measured from the full
microstate $\bsigma(t)$ (histograms) and
calculated from (\ref{eq:finalRSdistribution}) with only $m(t)$ and
$r(t)$ as input (dashed lines) are shown. The RS theory
 leading to the distribution (\ref{eq:finalRSdistribution}) turns out
to
give a good
qualitative description of the simulation data; significant
deviations are confined to the region above the AT line. Below the AT line
these
numerical results partially
justify {\em a posteriori} the ans\"{at}ze of self-averaging and
subshell equipartitioning, made to close the set of deterministic dynamical
 laws for the order parameters $m$ and $r$.

\subsection{Cooling in a Small External Field}

\begin{figure}
\vspace*{130mm}
\hbox to \hsize{\hspace*{5mm}\includegraphics{ps/waitingflow.ps}\hspace*{10mm}}
\vspace*{-50mm}
\caption{Flow in the $(m,r)$ plane of the order parameters $m(t)$ and
$r(t)$, at $T=0.1$ with a small external
field $\theta=0.1$. Initial state: $(m,r)=(0,0)$ (the paramagnetic
state). Fluctuating lines: three independent $N=3200$ simulations.
Smooth solid line: solution of RS flow equations.
Dashed lines: the $q=1$ line (upper) and the AT line (lower).}
\label{fig:coolingflow}
\end{figure}
\begin{figure}
\vspace*{9cm}
\hbox to \hsize{\hspace*{-45mm}\includegraphics{ps/waiting.ps}\hspace*{45mm}}
\vspace*{-1cm}
\caption{Evolution in time of the order parameters $m(t)$ (left
picture) and $r(t)$ (right picture), at $T=0.1$ with a small external
field $\theta=0.1$. Initial state: $(m,r)=(0,0)$ (the paramagnetic
state). Fluctuating lines: three independent $N=3200$ simulations.
Smooth line: solution of RS flow equations {\em below} AT line.
Dashed line: solution of RS flow equations {\em above} AT line.}
\label{fig:cooling}
\end{figure}

Next we study the evolution in time  of the order
parameters $m$ and $r$ that results after cooling the system
instantaneously
from $T=\infty$, the paramagnetic state
$(m,r)=(0,0)$, to $T=0.1$. For simplicity we choose $J_0=0$ and
$J=1$. An external field $\theta=0.1$ is applied in order
to obtain non-trivial evolution for the magnetisation (this field being small
assures the macroscopic state vector
eventually enters into the spin-glass region of the $(m,r)$ flow
diagram, above the AT line).

In figures \ref{fig:cooling} and \ref{fig:coolingflow} we compare the
result of performing numerical simulations (for an $N=3200$ system)
with the result of solving numerically the RS flow
equations (\ref{eq:mequation},\ref{eq:requation}).
At
least within the duration of the numerical experiments ($t\leq 10$
iterations/spin), the {\em
direction} of the flow in the $(m,r)$ plane is correctly described by
the RS flow equations, even above the AT line. Within the limitations
of our simulations the RS
theory, however, breaks down even before the AT line is crossed, in that the RS
flow
equations fail to describe an
overall slowing down of the macroscopic flow.

\subsection{Decay from a Fully Magnetized State}

\begin{figure}
\vspace*{130mm}
\hbox to \hsize{\hspace*{5mm}\includegraphics{ps/flowT00.ps}\hspace*{10mm}}
\vspace*{-50mm}
\caption{Flow in the $(m,r)$ plane of the order parameters $m(t)$ and
$r(t)$, at $T=0.0$ with $J_0=\theta=0$ and $J=1$ from
the fully magnetized initial state $(m,r)=(1,0)$.
Fluctuating lines: three independent $N=3200$ simulations.
Thick solid line: solution of RS flow equations.
Dashed lines: the $q=1$ line (upper) and the AT line (lower).}
\label{fig:kinzelflowT00}
\end{figure}
\begin{figure}
\vspace*{9cm}
\hbox to \hsize{\hspace*{-45mm}\includegraphics{ps/relaxT00.ps}\hspace*{45mm}}
\vspace*{-1cm}
\caption{Evolution in time of the order parameters $m(t)$ (left
picture) and $r(t)$ (right picture), at $T=0.0$ with
$J_0=\theta=0$ and $J=1$.
Fluctuating lines: three independent $N=3200$ simulations.
Smooth line: solution of RS flow equations {\em below} AT line.
Dashed line: solution of RS flow equations {\em above} AT line.}
\label{fig:kinzelrelaxT00}
\end{figure}

Finally we study the relaxation from the fully magnetized initial
state $(m,r)=(1,0)$ (a la Kinzel \cite{kinzel}, albeit for short
time-scales $t\leq 10$ only).
For simplicity we choose $J_0=\theta=0$ and
$J=1$. Figures \ref{fig:kinzelflowT00}
and \ref{fig:kinzelrelaxT00} show the
result of comparing  numerical simulations for an $N=3200$ system
with the result of solving numerically the RS flow
equations (\ref{eq:mequation},\ref{eq:requation}).
Again within the duration of the numerical experiments ($t\leq 10$
iterations/spin) the {\em
direction} of the flow in the $(m,r)$ plane is correctly described by
the RS flow equations, whereas the RS
theory apparently fails to describe the
overall slowing down that sets in even before the AT line is crossed
(which gives rise to the familiar remanent magnetisation
\cite{kinzel}).
In order to describe the slow relaxation of this remanent
magnetization above the AT line, measured rather in terms of a few thousand
iterations
per spin, we clearly need the RSB version of our theory.

\section{Discussion}

In this paper we have developed
a dynamical
theory, valid on finite time-scales, to describe
the Glauber dynamics of the SK model in terms of
deterministic flow equations for two macroscopic state variables: the
magnetisation and the spin-glass contribution to the energy.
Two transparant physical
assumptions, based on a systematic removal of microscopic memory
effects,  allow us to
calculate the time-dependent distribution of local aligment
fields in terms of the instantaneous prder parameters only and thereby obtain a
{\em closed}
set of flow equations for our two order parameters.
The theory produces  in a natural way
dynamical generalisations of the AT- and zero-entropy lines and of
Parisi's order parameter function $P(q)$. In equilibrium we recover
the standard results from equilibrium statistical mechanics,
including the full RSB equations.

In calculating the order parameter flow explicitly we have
made the replica-symmetric (RS) ansatz, as a natural first step.
A subsequent paper will be
devoted to the implications of breaking the replica symmetry (RSB).
We found that in
most of the flow diagram replica symmetry is stable.
Numerical simulations suggest that our
equations describe the shape of the local field distribution and
the macroscopic dynamics quite well  in the region where
replica symmetry is stable.
In the region of the flow diagram where the RS solution is unstable
 the flow {\em direction} as given by the RS theory seems still correct
 and the RS theory even predicts non-exponential
relaxation in the limit $T\rightarrow0$. However, the RS theory fails
to describe a rigorous slowing down
which, according to simulations, sets in near the de Almeida-Thouless
\cite{AT} line.
Intuitively one expects the breaking up of phase space,
 as indicated by the breaking of replica symmetry, to have a slowing down
effect on the macroscopic flow.
Preliminary investigations of the effect of replica symmetry breaking,
based on expansions just above the AT line and on a one-step symmetry
breaking a la Parisi, show that this is indeed
the case \cite{subsequent}.

The main appeal of our formalism we consider to be its
transparency.  The theory is formulated in terms of two directly
observable macroscopic state variables: the magnetisation and the
spin-glass contribution to the
energy per spin.
Furthermore the macroscopic laws are derived directly from
the underlying microscopic stochastic equations, given two key
assumptions.
An interesting
difference with existing (mostly Langevin) approaches is that in the present
formalism
replica theory
enters naturally as a mathematical tool in calculating the time-dependent
distribution of local alignment fields.
One of the two assumptions on which our analysis is based
(self-averaging of the macroscopic laws with respect to the frozen
disorder) is quite standard.
Both assumptions are supported by evidence from numerical
simulations. Based on the agreement between theory and simulations in
the RS region we believe that our two closure assumptions lead to a theory
which
captures the
main physics of the order parameter flow of the SK model on finite time-scales,
and that
 the impact of microscopic memory effects (which in the theory are
explicitely  removed) can be viewed, as in
\cite{coolensherrington,CF}, principally  as an
overall slowing down.
Our next step will be to investigate in detail
the RSB version of our dynamical laws, which will be the subject of a
subsequent paper.
\vsp\vsp

\noindent
{\bf Acknowledgements:}

\noindent
We would like to thank L. Cugliandolo and J. Kurchan for communicating
some of their
work prior to publication and S. Franz for interesting
discussions on SK dynamics.


\end{document}